\documentclass[psf,proceeding paper,accept,pdftex,moreauthors]{Definitions/mdpi} 
\firstpage{1} 
\pubvolume{41}
\issuenum{1}
\articlenumber{0}
\pubyear{2023}
\copyrightyear{2023}
\externaleditor{Academic Editor: Yue Zhao}
\datepublished{} 
\hreflink{https://doi.org/} 

\usepackage{url}
\Title{CE$\nu$NS Experiment Proposal at CSNS $^{\dagger}$}

\TitleCitation{CE$\nu$NS Experiment Proposal at CSNS}

\Author{Chenguang Su 
 $^{1}$, Qian Liu $^{1,*}$ and Tianjiao Liang $^{2,3}$ on behalf of the CLOVERS Collaboration}

\AuthorNames{Chenguang Su, Qian Liu and Tianjiao Liang on behalf of the CLOVERS Collaboration}

\AuthorCitation{Su, C.; Liu, Q.; Liang, T., on behalf of the CLOVERS Collaboration}

\address{%
$^{1}$ \quad School of Physical Sciences, University of Chinese Academy of Sciences, Beijing 100049, China; suchenguang17@mails.ucas.ac.cn
\\
$^{2}$ \quad Institute of High Energy Physics, Chinese Academy of Sciences, Beijing 100049, China; tjliang@ihep.ac.cn\\
$^{3}$ \quad Dongguan Institute of Neutron Science, Dongguan 523808, China\\
}

\corres{Correspondence: liuqian@ucas.ac.cn}

\firstnote{Presented at the 23rd International Workshop on Neutrinos from Accelerators, Salt Lake City, UT, USA, \mbox{30--31 July 2022}.} 

\abstract{The detection and cross-section measurement of Coherent Elastic Neutrino--Nucleus Scattering (CEvNS) are vital for particle physics, astrophysics, and nuclear physics. Therefore, a new CEvNS detection experiment is proposed in China. Undoped CsI crystals, each coupled with two Photon Multiplier Tubes (PMTs), will be cooled down to 77 K and placed at the China Spallation Neutron Source (CSNS) to detect the CEvNS signals produced by neutrinos from stopped pion decays occurring within the Tungsten target of CSNS. Owing to the extremely high light yield of pure CsI at 77 K, even though it only has a neutrino flux 60\% weaker than the COHERENT experiment, the detectable signal event rate is still expected to be 0.074/day/kg (0.053/day/kg for COHERENT). Low-radioactivity materials and devices will be used to construct the detector, and strong shielding will be applied to reduce the radioactive and neutron background. Dual-PMT readout should be able to reject PMT dark count background. Using all the strategies mentioned above, we hope to reach a 5.1$\sigma$ signal detection significance within six months of data collection with four 3kg CsI. This paper will discuss the experiment's design, as well as the estimation of the signal, various kinds of background, and expected signal sensitivity.} 


\begin{document}



\section{Introduction}

In general, the non-trivial interplay between neutrinos and individual nucleons, as well as the complex structure of the nucleus, makes a precise understanding of the interaction between neutrinos and the nucleus difficult to achieve. 
However, when the momentum transfer between a neutrino and a nucleus is small enough, which means the scattering would be elastic and the de Broglie wavelength corresponding to the transferred momentum much larger than the scale of the nucleus, from the point of view of the neutrino, the nucleons inside the nucleus hold almost the same position. Therefore, when calculating the cross section, the phases of scattering amplitudes contributed by different nucleons are almost the same, leading to a coherent enhancement of the cross section. This is the so-called coherent elastic neutrino--nucleus scattering (CE$\nu$NS) process.

The measurement of the CE$\nu$NS signal would benefit various aspects of physics. For particle physics, it provides a good way to measure the Weak-Mixing Angle in low momentum transfer circumstances and serves as another inspection of the Standard Model. For dark matter searching, a good knowledge of the CE$\nu$NS signal helps with the searching of WIMP dark matter candidates in which the CE$\nu$NS signals, as an important background, are hard to distinguish from real dark matter signals. For astrophysics, it points out a new way to detect solar neutrinos and supernova neutrinos, as their energy perfectly meets the coherent criterion. The accurate measurement of the cross section of the CE$\nu$NS process also helps to understand the outburst of core-collapsed supernovae whose energy is released mainly through neutrinos with energy of tens of MeV~\cite{ott2013core}. For nuclear physics, it can be used to measure the form factor of the nucleus when the exchanged momentum is large enough to introduce some non-coherent effect due to the structure of the nucleus, but also small enough to keep the scattering elastic. In addition, the CE$\nu$NS process has no threshold limit. Hence, the reactor neutrino spectrum under the 1.8MeV threshold of IBD process can be measured by the CE$\nu$NS process and the experiment's results could provide a good examination of nuclear physics models.

Since the measurement of the CE$\nu$NS process is vital for different fields of physics, many scientists have been devoted to this area since the first theoretical prediction was published by Daniel Z. Freedman in 1974~\cite{freedman1974coherent}. Owing to the coherent enhancement, its cross section is approximately proportional to the square of the neutron number in the nucleus~\cite{drukier1984principles}, making it much larger than any other neutrino--matter interactions. However, the signal produced by the recoiled nucleus is so weak that the CE$\nu$NS signal had not been observed until 2017 by the COHERENT collaboration using $\pi^{+}$ decay-at-rest neutrinos from the Spallation Neutron Source (SNS) in Oak Ridge~\cite{akimov2017observation}. There are also many other groups trying to detect CE$\nu$NS signals using reactor neutrinos with various kinds of technologies applied. For instance, a cryogenic superconductor calorimeter was selected by NUCLEUS in France~\cite{angloher2019exploring} while an astronomical CCD was applied by CONNIE in Mexico~\cite{aguilar2019exploring}.

Despite the significant effort that has been put into the searching of CE$\nu$NS signal, independent CE$\nu$NS signal detection verification still remains a blank to fill. Here, we propose the CLOVERS experiment: Coherent eLastic neutrinO(V)-nucleus scattERing at China Spallation Neutron Source (CSNS). The CSNS provides neutrinos with almost the same spectrum as SNS in Oak Ridge. 
The experiment design and the estimation of signal and background are discussed in Section \ref{Exp_Deg} and Section \ref{Sig_Bkg_Esti}. A sensitivity estimation and the experiment schedule are presented in Section \ref{Sens} and Section \ref{Sched}.


\section{Experiment Design}\label{Exp_Deg}

The common difficulties faced during neutrino detection experiments are the low cross section, weak signal, and high background. The CE$\nu$NS experiment shares the latter two challenges, while the first one is alleviated by the coherent enhancement of the cross section. The observable energy generated by the recoiled nucleus is only several keV, which requires the detector's threshold to be very low. Since the signals are weak, the background must be strongly suppressed to ensure that the signals are not overwhelmed. Therefore, an optimized shielding structure is necessary. The following part of this section describes our selection of the neutrino source and our design of the detector and shielding structure.

\subsection{Selection of Neutrino Source}\label{Neu_Soc}

\begin{figure}[H]
\includegraphics[width=10.5 cm]{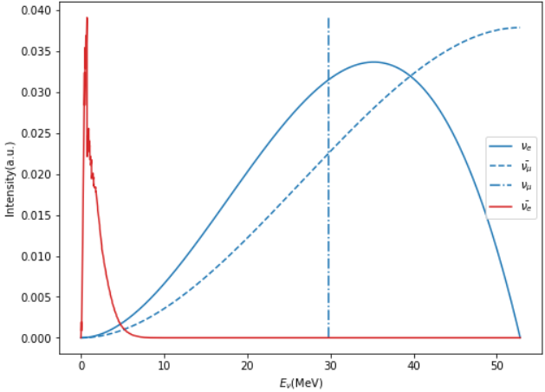}
\caption{Neutrino energy spectra of reactors (\textbf{red}) and spallation neutron source (\textbf{blue}). The energy neutrinos from spallation neutron  are significantly higher reactor neutrinos~\cite{Avignone2002kc, riyana2019calculation}. \label{fig_Nu_Spec}}
\end{figure}

The China Spallation Neutron Source (CSNS) has been selected as our neutrino source. It is located in Dongguan, Guangdong province, in China. At CSNS, a beam of protons is accelerated to $1.6$ GeV and impinges on a Tungsten target with a repetition rate of $25$~Hz. Neutrinos, including $\nu_{\mu}$, $\bar{\nu_{\mu}}$, and $\nu_{e}$, are generated by the decay of target-stopped $\pi^{+}$ resulting from the proton impingement. As a result, the neutrinos are highly pulsed, which is advantageous for suppressing background signals that are evenly distributed in time. The energy of neutrinos produced by $\pi^{+}$ decay-at-rest primarily ranges between $20$ and $50$~MeV, which is approximately one order of magnitude higher than that of reactor neutrinos (Figure~\ref{fig_Nu_Spec}). This higher energy range makes the detection of CE$\nu$NS signals much~easier.

\begin{figure}[H]
\includegraphics[width=12 cm]{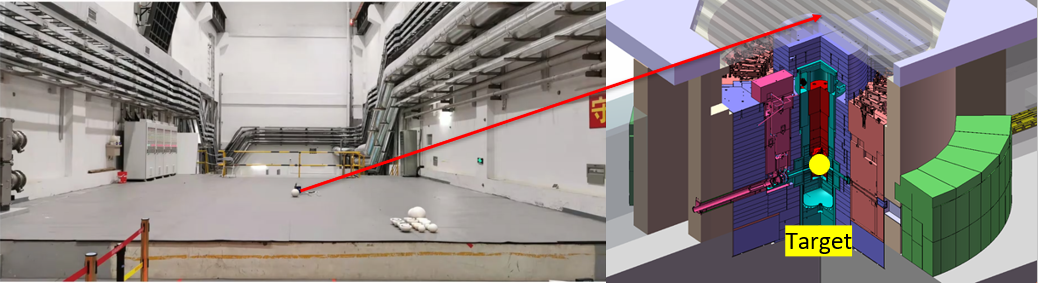}
\caption{Scene picture of the platform (\textbf{left}) and its relative postion in CSNS (\textbf{right}). The platfrom is $8.2$ m right above the Tungsten target. \label{fig_Loc}}
\end{figure}

Currently, CSNS is operating with a beam power of $140$ kW. According to simulations performed using FLUKA, the neutrino production rate is estimated to be approximately $0.17$ neutrinos per proton per flavor at CSNS~\cite{HuangMY2016}. Our detector will be placed on a platform located {$7.7$} m above the target. Considering the thickness of the shielding structure and the detector encapsulation to be {$2.8$} m, the neutrino flux at the detector's location is calculated to be 2.42 $\times$ 10$^{10}$/cm$^{2}$/h/falvor 
. This corresponds to approximately $60\%$ of the flux observed in the COHERENT experiment~\cite{scholz2018first}. Figure \ref{fig_Loc} shows the platform (left) and its relative position within CSNS (right).

\unskip

\subsection{Detector Design}\label{Det_Deg}

\begin{figure}[H]
\includegraphics[width=10.5 cm]{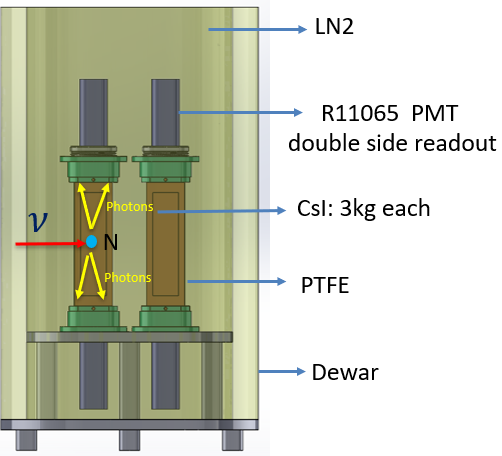}
\caption{A schematic of the detector. The detector contains four sub-detectors in a Dewar filled with liquid nitrogen. Each sub-detector is composed of one 3 kg undoped CsI and two R11065 PMTs. \label{fig_DetSche}}
\end{figure}

To achieve a recoil energy threshold of approximately $1$ keV, a detector as shown in Figure \ref{fig_DetSche} has been designed. The detector consists of several sub-detectors housed in a large Dewar. Each sub-detector is composed of a PTFE fabrication shell, two R11065 Hamamatsu photomultiplier tubes (PMTs), and one $3$ kg undoped cesium iodide (CsI) crystal. The Dewar will be filled with liquid nitrogen to immerse the sub-detectors and maintain a stable cryogenic temperature of $77$ K. While the light yield of undoped CsI is lower than that of CsI(Na) and CsI(Tl) at room temperature~\cite{AMSLER2002494, Woody106667}, it increases by more than 15 times when cooled down to $77$ K~\cite{AMSLER2002494, Mikhailik_CsI}. When coupled with R11065 PMTs, the light yield of undoped CsI has been reported to reach $33.5$PE/keV$_{\text{ee}}$~\cite{Ding2020uxu}, more than twice that of CsI(Na) measured by the COHERENT collaboration at room temperature~\cite{scholz2018first}. The high light yield enables us to lower the threshold.

The two PMTs in each sub-detector form a coincidence system to reject backgrounds generated by a PMT dark count background, including electron emission on the cathode or dynodes, and Cherenkov light generated by charged particles passing through the PMT window. This background dominates in the COHERENT CsI(Na) experiment~\cite{scholz2018first} in the few photoelectron (PE) region. Since the PMT dark count background is independent from one PMT to another, this background can be suppressed by three orders of magnitude by applying a two PMT readout coincidence system, which requires at least one photoelectron to be detected in each PMT. The suppression effect is discussed in detail in Section \ref{est_bkg}.

The four sub-detectors also form an anti-coincidence system. The probability of one particle producing signals in more than one sub-detector is negligible for neutrinos but much larger for fast neutrons and $\gamma$ rays. Events in which scintillation signals are observed in more than one sub-detector will be regarded as background. This strategy can significantly reduce the fast neutron background. The details of the effect of this strategy are discussed in Section \ref{est_bkg}.
\vspace{-6pt}

\subsection{Shielding Structure}\label{Shielding}

\begin{figure}[H]
\includegraphics[width=10.5 cm]{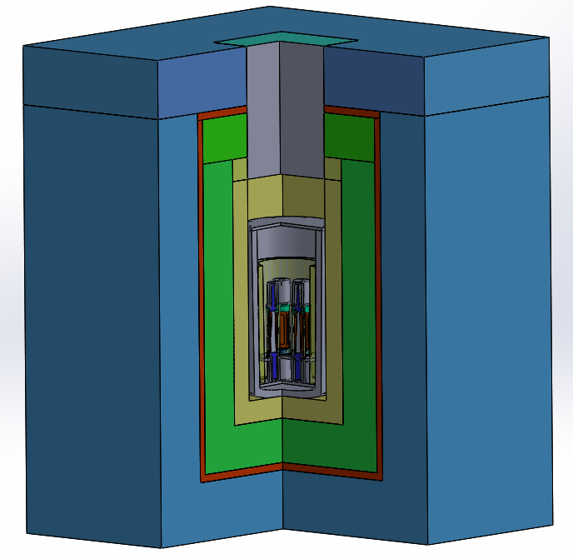}
\caption{A schematic of the preliminary shielding structure design. The shielding components from the inside out are as follows. (\textbf{{1}
}) Yellow green: $5$ cm OHFC. (\textbf{{2}}) Gray: Dewar. (\textbf{{3}}) Yellow: $30$ cm HDPE. (\textbf{{4}}) Green: $60$ cm Lead. (\textbf{{5}}) Red: $5$ cm $\mu$ veto plastic scintillator. (\textbf{{6}}) Blue: $80$ cm HDPE. \label{fig_Shield_Sche}}
\end{figure}

In order to achieve a $5\upsigma$ detection of the CE$\nu$NS signal in approximately one year, the background event rate needs to be reduced to the same magnitude as the CE$\nu$NS signal. To achieve this goal, a preliminary shielding structure as shown in Figure \ref{fig_Shield_Sche} has been designed. Inside the Dewar (gray), the four sub-detectors are surrounded by a $5$ cm-thick layer of OHFC (oxygen-free high-conductivity copper) to shield against the radioactive background produced by the stainless steel in the Dewar and the inner layer shielding materials. Outside the Dewar, there is a $30$ cm thick layer of HDPE (high-density polyethylene), followed by a $60$ cm-thick layer of lead. The lead shield aims to reduce the $\gamma$-ray background, while the innermost layer of HDPE is designed to slow down and stop fast neutrons produced by high-energy neutrons (>50 MeV) interacting with lead nuclei. The lead shield is encased by a $5$ cm-thick $\mu$ veto plastic scintillator to tag cosmic ray events. The outermost layer consists of $80$ cm-thick HDPE, which serves as a strong moderator for fast neutrons. The total thickness of HDPE in this design reaches $1.1$ m because the fast neutrons escaping from the Tungsten target are expected to be the main background on the platform. A detailed discussion of the fast neutron background is provided in Section \ref{est_bkg}.

It is important to note that this shielding structure is preliminary. As the on-site background measurement on the platform at CSNS is currently underway, this design will be adjusted and optimized based on our measurement results.

\subsection{Data Taking Strategy and Event Selection}

A data-taking strategy is proposed, taking into consideration the characteristics of the neutrino source and the detector. The strategy is outlined as follows:

\begin{enumerate}
\item The $25$ Hz proton beam trigger signal provided by CSNS will be used as an external trigger for the experiment's data collection, suppressing steady-state backgrounds such as cosmic rays and environmental noise by four orders of magnitude. Each trigger corresponds to one event, which will be referred to later.
\item A complete waveform signal from each PMT will be recorded by a flash ADC with a sampling rate of $1$ GHz. Each waveform will extend for 50 $\upmu$s, with a 10 $\upmu$s signal region and a 40 $\upmu$s pretrace. Offline waveform analysis will be applied to extract CE$\nu$NS signal candidates.
\item Every event will be recorded with a time tag and a $\mu$ veto tag. By referencing the time tag to the beam power monitor of CSNS, fluctuations in the proton beam power can be neutralized. The $\mu$ veto tag signal from the $\mu$ veto system will also be used to reject events that may be contaminated by cosmic rays.
\end{enumerate}

Although the $25$ Hz proton beam trigger can reduce the steady-state background by four orders of magnitude, certain event selection criteria are still required to further reduce the background to meet the standard mentioned in Section \ref{Shielding}. The criteria are listed as~follows:

\begin{enumerate}
\item The event must not be tagged by the $\mu$ veto system in order to reject events that may be contaminated by cosmic rays.
\item For each waveform, the number of photoelectrons (NPE) found in the pretrace region should be smaller than $3$ to suppress the afterglow background introduced by other particles hitting the CsI detector just a few microseconds before the trigger.
\item For each sub-detector, at least one PE should be detected in both PMTs. This criterion aims to reduce the PMT dark count background, which is independent from one PMT to another.
\item For the entire detector system, events with more than one sub-detector satisfying criterion 3 would be excluded, as neutrons and $\gamma$ rays are much more likely to produce signals in different sub-detectors.
\end{enumerate}

These event selection criteria, in combination with the proton beam trigger and $\mu$ veto system, will help further reduce the background and enhance the sensitivity of the CE$\nu$NS signal detection.


The selection efficiencies of all criteria for CE$\nu$NS signals have been investigated. However, the efficiencies of the first two criteria need to be determined on-site once the entire detector and shielding system have been constructed; these are currently not ready. Hence, the efficiencies of similar cuts applied by the COHERENT collaboration are adopted in the sensitivity estimation in Section \ref{Sens}~\cite{scholz2018first}. The adopted efficiencies are $98.9\%$ for criterion 1 and $73.8\%$ for criterion 2. Considering that the cosmic ray levels and other steady-state backgrounds responsible for most afterglow events should be similar in CSNS and SNS, while the afterglow of undoped CsI at $77$ K is much weaker than that of CsI(Na) at room temperature, these estimations of the efficiencies for the first two criteria should be conservative overall.

The efficiency of criterion 3 is considered by assuming that the probabilities of scintillation photons being detected by the two PMTs are equal on average, which is reasonable enough considering the longitudinal symmetry of the detector. An analytical calculation based on the binomial distribution has been carried out to estimate the selection efficiency of scintillation signals for different numbers of detected photoelectrons (NPE). The efficiency of criterion 4 is considered to be $100\%$, since the possibility of a single neutrino producing signals in different sub-detectors is negligible.

\begin{figure}[H]
\includegraphics[width=10.5 cm]{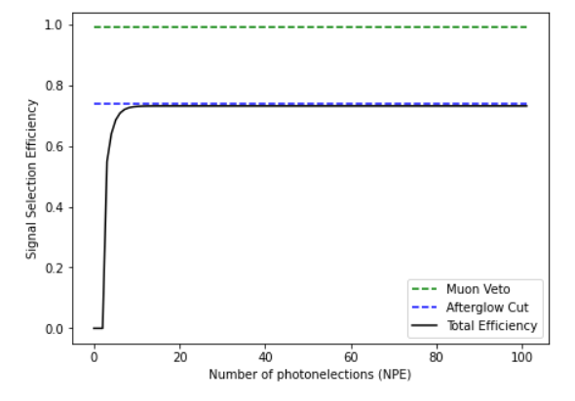}
\caption{Efficiency of event selection criteria. \textbf{Green}: Efficiency of $\mu$ veto cut \cite{scholz2018first}. \textbf{Blue}: Efficiency of afterglow cut~\cite{scholz2018first}. \textbf{Black}: Total Efficiency. \label{fig_Eff}}
\end{figure}

Figure \ref{fig_Eff} shows the estimated CE$\nu$NS signal selection efficiency for different numbers of detected NPE. The green and blue lines represent the efficiencies of the $\mu$ veto and afterglow cuts, respectively, as taken from~\cite{scholz2018first}. Both cuts have constant efficiencies for all events. The black line represents the total efficiency of all criteria obtained by multiplying the efficiency of each cut together. Criterion 3 contributes to the rising shape of the curve, and its efficiency is equal to the efficiency for all scintillation events, including signals generated by particles depositing energy in CsI.
\vspace{-6pt}

\section{Estimation of CE$\nu$NS Signal and Background}\label{Sig_Bkg_Esti}

With the experiment design described above, the CE$\nu$NS signal and background in this experiment can be estimated. In this analysis, the detector is assumed to be composed of four $3$ kg undoped CsI sub-detectors, totaling $12$ kg of undoped CsI. The data-collection time is assumed as half a year. Based on the results obtained in this analysis, the expected sensitivity of this experiment can be determined.

\subsection{Estimation of CE$\nu$NS Signal}\label{est_sig}

The estimation of the CE$\nu$NS signal involves two steps. First, we need to consider the CE$\nu$NS recoiled energy distribution of nuclear recoils induced by neutrinos. Second, we need to take into account the energy response of the detector to recoiled nuclei in order to obtain the expected spectrum of NPE detected by the PMTs.

\begin{figure}[H]
\includegraphics[width=10.5 cm]{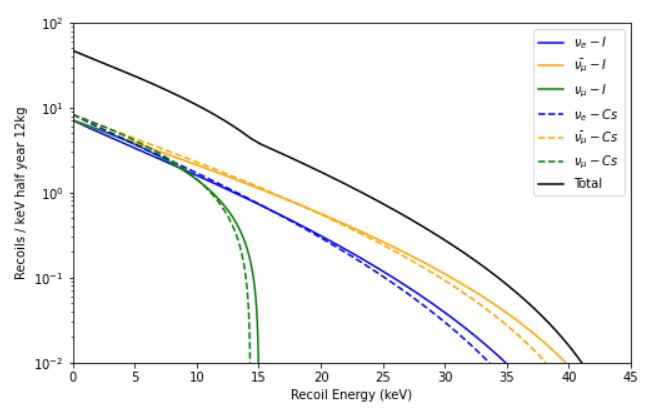}
\caption{Expected recoiled energy distribution of CE$\nu$NS interaction detected by a $12$ kg cryogenic undoped CsI detector $10.5$ m away from the Tungsten target with a half-year of data collection. The contributions from different flavors of neutrinos and different isotopes are also shown. \label{fig_CEvNS_Recoil}}
\end{figure}

The expected CE$\nu$NS recoiled energy distribution can be calculated numerically by considering the neutrino flux, detector mass, and the CE$\nu$NS differential cross section~\cite{drukier1984principles}. Figure \ref{fig_CEvNS_Recoil} shows the calculated result, including the total event rate distribution and the event rates for the three different neutrinos generated in CSNS. The total CE$\nu$NS event rate reaches $303$ events per half year per $12$ kg of detector mass, equivalent to $0.14$ events per day per kg.
\vspace{-6pt}

\begin{figure}[H]
\includegraphics[width=10.5 cm]{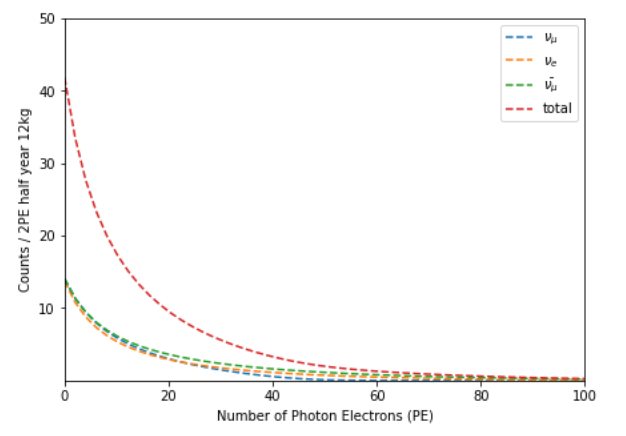}
\caption{Expected NPE spectra of CE$\nu$NS signal. Contributions from different flavors of neutrinos are also shown. \label{fig_CEvNS_NPE}}
\end{figure}

The energy response of the detector to recoiled nuclei involves two steps. First, the light yield with respect to the energy deposited through the ionization process, which is often calibrated using $\gamma$ rays or $\beta$ rays. Second, the quenching factor (QF) of the detector material, which represents the efficiency of nuclear recoil energy transformation into ionization energy. For this estimation, a light yield of $33.5$PE/ keV$_{\text{ee}}$ for undoped CsI at $77$ K~\cite{Ding2020uxu} is adopted. The quenching factor of undoped CsI is based on measurements conducted by the COHERENT collaboration~\cite{COHERENT2021pcd}.  By convolving the recoiled energy distribution with the energy response of the detector, Figure \ref{fig_CEvNS_NPE} is obtained, which shows the expected spectrum of photoelectrons detected by the PMTs. 

\subsection{Estimation of Background}\label{est_bkg}
The background in the CE$\nu$NS experiment arises from various sources. (i) Beam-related neutrons (BRN) are produced when protons impinge on the Tungsten target. Although a shield consisting of {$6.7$} m of steel and $1$ m of concrete is placed between the target and the platform, some fast neutrons can still escape and reach the platform. These fast neutrons can generate nuclear recoil signals by scattering with Cs and I nuclei, which are indistinguishable from real CE$\nu$NS signals and cannot be reduced by the proton beam trigger. (ii) The PMT dark count events can occur within the signal region of the recorded waveform. Since most CE$\nu$NS signals and most PMT dark count signals generate only a few detectable photoelectrons (PEs), the PMT dark count background could be significant in the low NPE region. (iii) The materials and devices used to construct the detector and shield unavoidably contain some long-lived radioactive isotopes. Their decay can introduce $\gamma$ and $\beta$ background in the detector. (iv) The environmental $\gamma$ background, resulting from the decay of long-lived radioactive isotopes in rock and building materials, is present everywhere, including the platform.

To evaluate the influence of these backgrounds on the experiment, a simulation software framework based on Geant4 has been developed. This framework allows for a detailed investigation of the different types of background. All the results presented in the following sections are obtained assuming a detector consisting of four $3$ kg undoped CsI sub-detectors and a data-collection time of half a year.

\subsubsection{Beam Related Neutron Background}\label{BRN}

\begin{figure}[H]
\includegraphics[width=12 cm]{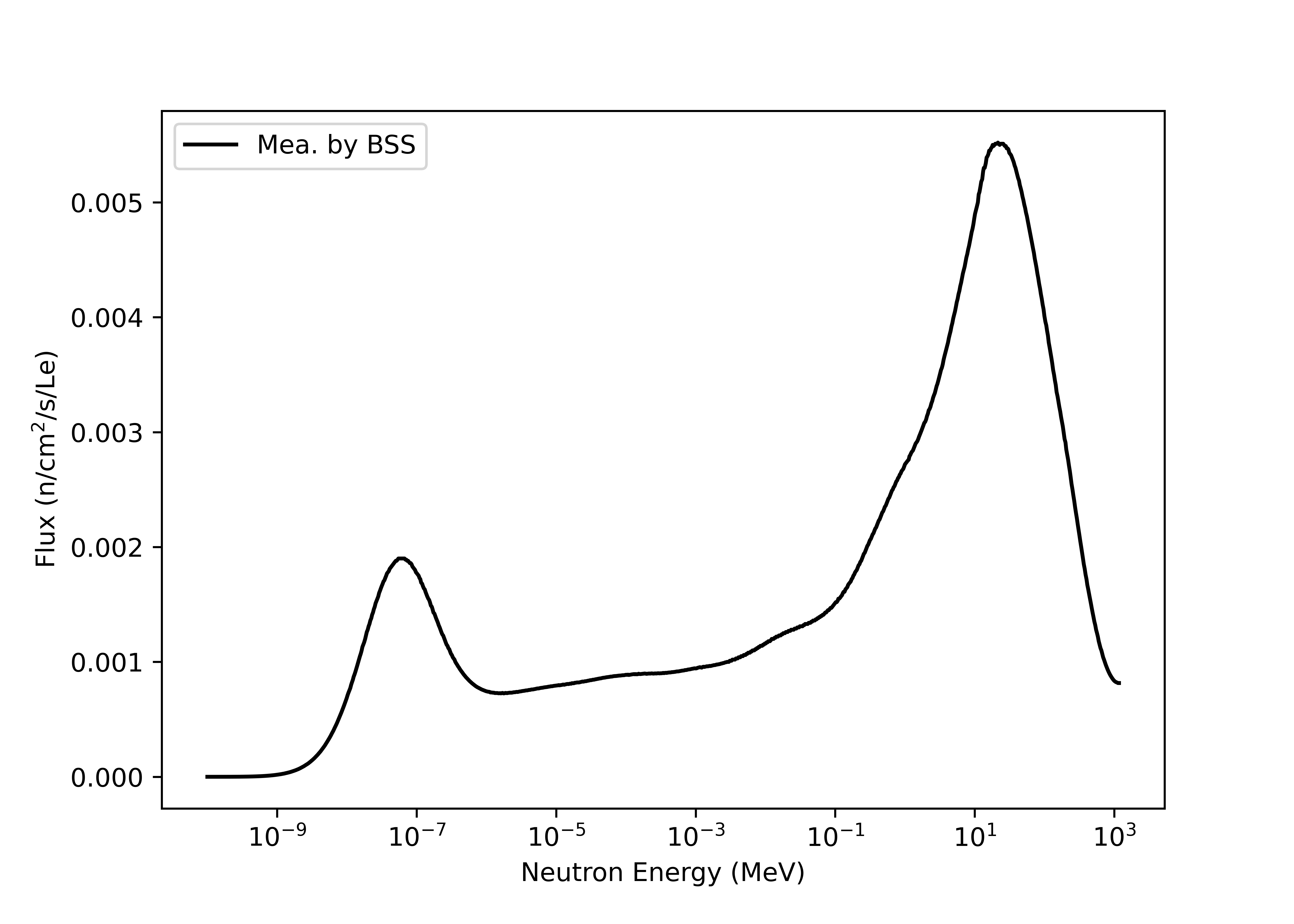}
\caption{Neutron spectrum measured by a ${}^{3}$He multi-sphere neutron spectrometer. 
\label{fig_Neutron_Spec}} 
\end{figure}

\begin{figure}[H]
\includegraphics[width=10.5 cm]{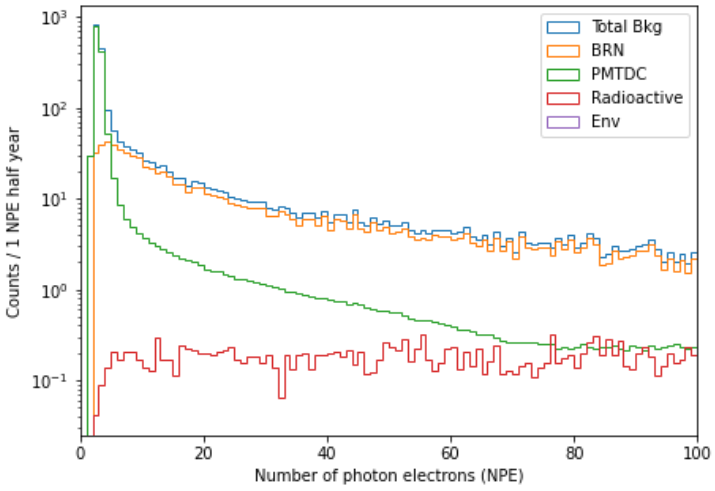}
\caption{The summary of contributions of background from different sources after all cuts applied. PMT dark count background (\textbf{green}) dominates in the low NPE region, while the BRN background (\textbf{orange}) prevails in the high NPE region. The radioactive background (\textbf{red}) contributes a low-flat component, and the environmental background (\textbf{purple}) is too weak to be seen. Four $3$ kg undoped CsI sub-detectors are considered.\label{fig_AllBkg}}
\end{figure}

A ${}^3$He multi-sphere neutron spectrometer was utilized to measure the neutron spectrum on the platform and outside the facility. Figure \ref{fig_Neutron_Spec} presents the unfolded neutron spectrum on the platform, obtained using a method similar to the one described in~\cite{Li2022cjw_Bonar_Spec}. The integrated neutron flux from Figure \ref{fig_Neutron_Spec} is $4.8 \times10^{-2}$ n/cm$^2$/s, which is approximately one order of magnitude higher than the flux measured outside the facility, which is \mbox{$5.3 \times10^{-3}$ n/cm$^2$/s}. Using this neutron spectrum as input and considering the entire shielding structure, a simulation was conducted to assess the amount of neutron background generated. After applying all the event selection criteria, the surviving neutron background spectrum is shown as the orange line in Figure \ref{fig_AllBkg}
. It is important to note that according to the simulation, event selection criterion 4 can reject over 70\% of the neutron~background.
\subsubsection{PMT Dark Count Background}\label{PMTDC}

\begin{figure}[H]
\includegraphics[width=10.7 cm]{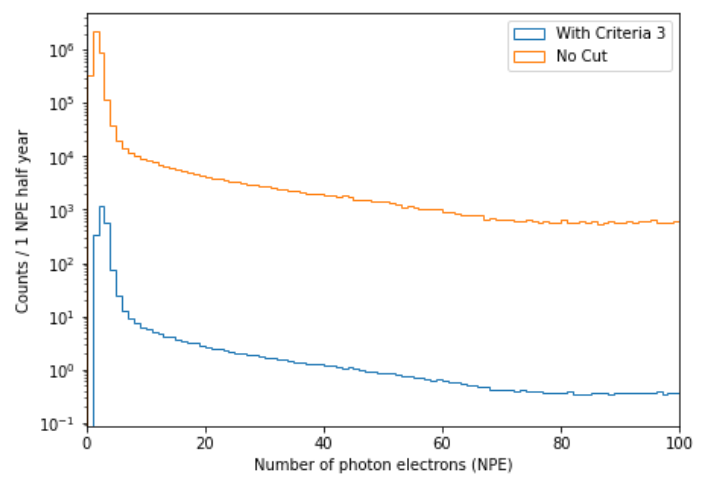}
\caption{PMT dark count spectra with (\textbf{blue}) and without (\textbf{orange}) applying event selection criteria 3. Four couples of PMTs of four sub-detectors are considered. Event selection criteria 3 can suppress this background by three magnitudes. \label{fig_Cr3_PMTDC}}
\end{figure}

A PMT dark count spectrum at $77$ K was obtained by setting the data acquisition system to a self-trigger mode and adjusting the threshold to trigger single-photon electron (SPE) signals. The dark count rate was measured to be an average of $111$ Hz and remained stable over a $24$-hour data-collection period. To investigate the effect of criterion 3 on this background, a toy Monte Carlo analysis was performed. Four pairs of PMTs from the four sub-detectors were considered. Figure \ref{fig_Cr3_PMTDC} illustrates the background level with and without the application of event selection criterion 3. The PMT dark count background can be suppressed by three orders of magnitude by this criterion.

\subsubsection{Long-Lived Radioactive Isotopes Background}\label{Radioac}

\begin{figure}[H]
\includegraphics[width=10.5 cm]{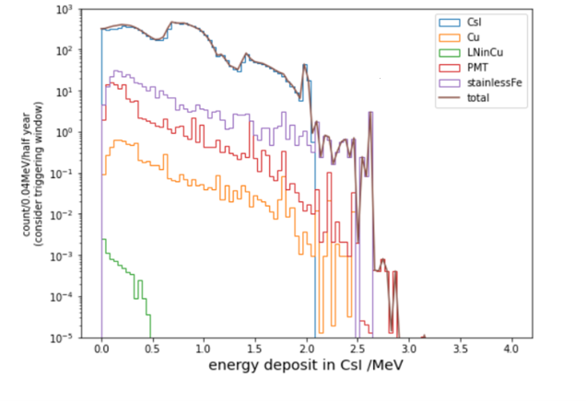}
\caption{The spectra of energy deposited in CsI from radioactive isotopes in different materials and devices. Four $3$ kg undoped CsI sub-detectors are assumed. The background from CsI (\textbf{blue}) dominates, followed by background from stainless steel of the Dewar (\textbf{purple}),  PMTs (\textbf{red}) and Copper (\textbf{orange}). The contribution from liquid nitrogen (\textbf{green}) is very small. 
	\label{fig_RadioactiveBkg}}
\end{figure}

The concentrations of different long-lived radioactive isotopes in various materials and devices are provided in Table \ref{table_radioac}. The impact of the decay radiation from these isotopes was simulated, taking into account their decay chains and assuming a state of decay equilibrium. Figure \ref{fig_RadioactiveBkg} illustrates the energy deposition spectrum in CsI resulting from the radioactive backgrounds of different materials and devices. It is evident that the radioactive background from the CsI crystal itself dominates.
\vspace{-6pt}

\unskip
\begin{table}[H]
\caption{The concentration
 of radioactive isotopes in different materials and devices.\label{table_radioac}}
	\begin{adjustwidth}{-\extralength}{0cm}
		\newcolumntype{C}{>{\centering\arraybackslash}X}
		\setlength{\tabcolsep}{1.5mm}
		\begin{tabularx}{\fulllength}{@{}CCCCCCCCC@{}}
			\toprule
			               & \textbf{PTFE} & \textbf{Fe} & \textbf{HDPE} & \textbf{PMT} & \textbf{Lead} & \textbf{LN2} \textbf{\textsuperscript{1}} & \textbf{CsI}                & \textbf{OFHC}              \\ \midrule
\textbf{K40 
}   & 0.343         & 60          & 43.3          & 37.1         & -             & -            & -                           & -                          \\
\textbf{Ra226} & 0.12          & -           & -             & -            & 3             & -            & -                           & -                          \\
\textbf{Ra228} & 0.11          & -           & -             & -            & -             & -            & -                           & -                          \\
\textbf{Th228} & 0.065         & -           & -             & -            & 1             & -            & -                           & -                          \\
\textbf{U238}  & 1.96          & -           & 19.8          & 5.2          & 0.12          & -            & -                           & 0.077                      \\
\textbf{Ac228} & -             & 70          & -             & -            & -             & -            & -                           & -                          \\
\textbf{Bi214} & -             & 25          & -             & -            & -             & -            & -                           & -                          \\
\textbf{Pb212} & -             & 70          & -             & -            & -             & -            & -                           & -                          \\ 
\textbf{Pb214}          & -             & 25          & -             & -            & -             & -            & -                           & -                          \\
\textbf{Th234}           & -             & 200         & -             & -            & -             & -            & -                           & -                          \\
\textbf{Tl208}          & -             & 70          & -             & -            & -             & -            & -                           & -                          \\
\textbf{Th232}          & -             & -           & 12.2          & 13.4         & -             & -            & -                           & 0.005                      \\
\textbf{Pb210}          & -             & -           & -             & -            & 240,000        & -            & -                           & -                          \\
\textbf{Ar39}           & -             & -           & -             & -            & -             & 0.01         & -                           & -                          \\
\textbf{Co60}           & -             & 17          & -             & -            & -             & -            & -                           & -                          \\
\textbf{Cs137}          & 0.17          & 6           & -             & -            & -             & -            & 150 & -  \\
\textbf{Cs134}          & -             & -           & -             & -            & -             & -            & 50  & - \\\midrule
\textbf{Unit}           & mBq/kg        & mBq/kg      & mBq/kg        & mBq/PMT      & mBq/kg        & mBq/kg       & mBq/kg                      & mBq/kg                     \\\midrule
\textbf{Reference}      & Xenon1T~\cite{XENON2017fdb_Bkg_PTFE_Cu_HDPE}       & ILIAS ANAIS~\cite{FeBkg} & Xenon1T~\cite{XENON2017fdb_Bkg_PTFE_Cu_HDPE}       & Taishan LAr~\cite{Wei2020rbm} & EDELWEISS \cite{EDELWEISS2013wrh_Bkg_Lead}    & Taishan LAr~\cite{Wei2020rbm}  & KIMS~\cite{Kim2003ms_CsIBkg}                        & Xenon1T~\cite{XENON2017fdb_Bkg_PTFE_Cu_HDPE}\\
     \bottomrule
		\end{tabularx}
	\end{adjustwidth}
	\noindent{\footnotesize{\textsuperscript{1} The background of liquid nitrogen is estimated by assuming the nitrogen reaches a purity of 99.99\% and the remaining impurities are all argon. The radioactive background level of atmosphere argon is taken from~\cite{Wei2020rbm}}.}
\end{table}
\subsubsection{Environmental $\gamma$ Background}\label{Env}

\begin{figure}[H]
\includegraphics[width=10.5 cm]{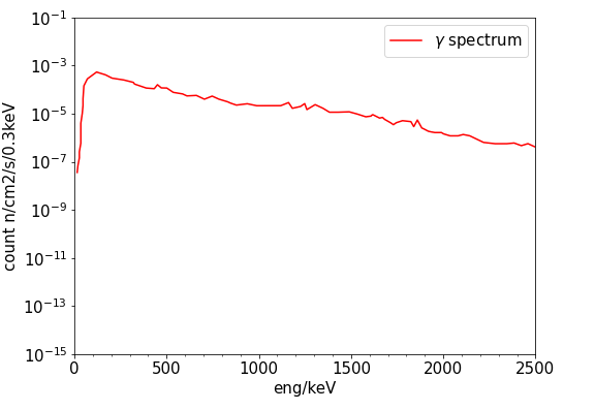}
\caption{The environmental $\gamma$ background redraw from the measurement by CDEX collaboration in CJPL~\cite{Ma2020rpd_CDEX_Gamma}. \label{fig_Env}}
\end{figure}

The contribution of environmental $\gamma$ background was estimated using the spectrum shown in Figure \ref{fig_Env} as input. This spectrum was measured by the CDEX collaboration in CJPL (China Jinping Underground Laboratory)~\cite{Ma2020rpd_CDEX_Gamma}. It should be noted that an underground environment such as CJPL is expected to have stronger radioactivity from ${}^{222}$Rn and other radioactive isotopes in the surrounding rock, making this estimation conservative.
\vspace{-6pt}

According to our simulation, owing to the strong shielding effect of lead against $\gamma$ radiation, the number of $\gamma$ rays capable of penetrating the shielding and depositing energy in the CsI detector is extremely low. Consequently, the contribution of environmental $\gamma$ background is negligible and not visible in Figure \ref{fig_AllBkg}.

\subsubsection{Summary of Background}

After applying all the event selection criteria, Figure \ref{fig_AllBkg} provides a summary of the contributions of background sources that survive all cuts. In the region of NPE smaller than five, the dominant background contribution comes from PMT dark counts. In the higher NPE region, the background from beam-related neutrons (BRN) becomes more prevalent. The radioactive background contributes a relatively low and flat component, while the environmental background is too weak to be visible.

Indeed, there are other possible background sources such as neutrino-induced neutrons (NIN) and cosmic ray-induced short-lived radioactive isotopes (CRSRI). However, these backgrounds have been found to be negligible by the COHERENT collaboration~\cite{scholz2018first}. The event rates of both NIN and CRSRI depend on the specific shielding structure. The NIN event rate also relies on the neutrino flux, while the CRSRI event rate is proportional to the cosmic ray rate. In our estimation, we neglect the contributions from NIN and CRSRI. This is based on the fact that our experiment shares a similar neutrino flux, shielding structure, and overburden to cosmic rays with the COHERENT experiment, with differences within one order of magnitude.



\section{Expected Sensitivity}\label{Sens}

\begin{figure}[H]
\includegraphics[width=10.5 cm]{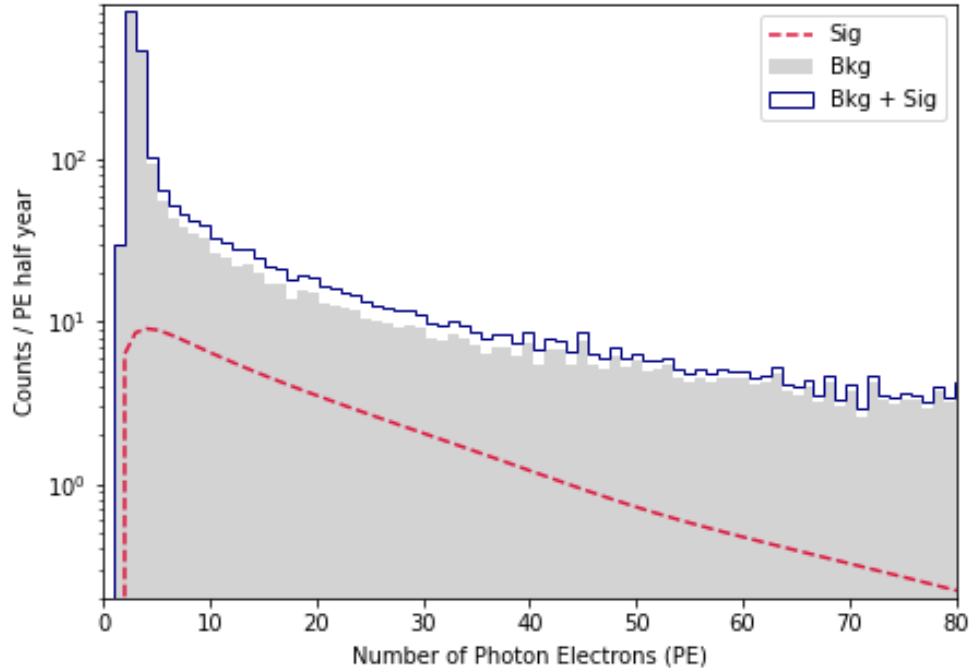}
\caption{The expected spectra of CE$\nu$NS events (\textbf{dashed red}), background events (\textbf{shadowed gray}) and their summation (\textbf{solid blue}) with all event selection criteria applied. Four $3$ kg undoped CsI sub-detectors are considered. The first few bins of background and the summation reach out of the y-axis range because the PMT dark count background is very high in this region. \label{fig_Sens_Esti}}
\end{figure}

Using the estimated signal and background spectra mentioned above, the expected sensitivity of this experiment can be evaluated. Figure \ref{fig_Sens_Esti} depicts the expected spectra of CE$\nu$NS events, background events, and their summation with all event selection criteria applied. This evaluation assumes a detector consisting of four $3$ kg undoped CsI sub-detectors and a data-collection time of half a year.
A signal region between $4$ and $72$ NPE is selected to maximize the signal-to-background ratio. The lower limit of $4$ NPE is chosen to exclude most of the PMT dark count background, corresponding to a detection threshold of approximately $1.5$ keV$_{\text{nr}}$ for nuclear recoils. Within this signal region, the total detectable signal event rate is $0.074$/day/kg, {which is higher than the $0.053$/day/kg event rate estimated by COHERENT}~\cite{akimov2017observation}. The total background event rate is $0.386$/day/kg, equivalent to $160$/half year/12 kg and $833$/half year/12 kg, respectively. The composition of the background events is listed in Table \ref{table_bkg_afcut}.

\begin{figure}[H]
\includegraphics[width=10.5 cm]{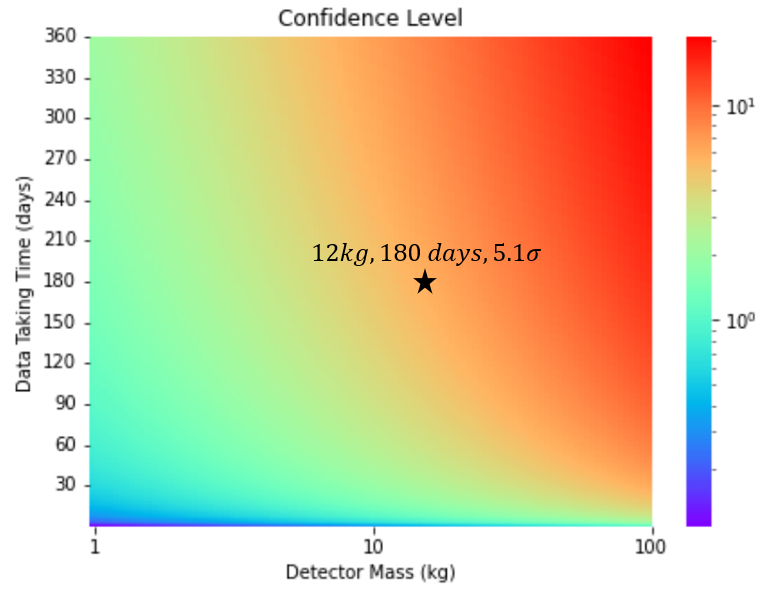}
\caption{The expected confidence level varying with different detector mass and data-collection time. If a $12$ kg CsI detector is employed and collecting data for half an year (180 days), the C.L. can reach $5.1\sigma$ (the pentagram).  \label{fig_CL_m_t}}
\end{figure}

The expected confidence level (C.L.) of this experiment is calculated using the following formula:
\begin{linenomath}
\begin{equation}
C.L. = \frac{N_{sig}}{\sqrt{N_{sig} + N_{bkg}}},
\end{equation}
\end{linenomath}
where $N_{\text{sig}}$ is the expected event number of CE$\nu$NS signal and $N_{\text{bkg}}$ is the expected event number of the background. Based on the experiment setup assumed above, the C.L. is expected to reach $5.1\sigma$ in half a year. Figure \ref{fig_CL_m_t} illustrates the expected C.L. varying with different detector mass and data-collection time.
{Systematic uncertainties of this experiment mainly arise from the uncertainties of the neutrino flux and the quenching factor of CsI. However, these uncertainties can be reduced to 2--3\% through dedicated experimental measurements, making them small enough compared to the statistical uncertainties}~\cite{COHERENT2021pcd, COHERENT:2021xhx}. {Therefore, the systematic uncertainties are not considered in this preliminary sensitivity estimation.}

Please note that the contribution to the C.L. from the arrival time profile of events with respect to the proton beam trigger has not been included in this estimation yet. Since the arrival time profile of CE$\nu$NS signals is highly correlated with that of the proton beam, it significantly differs from the arrival time profiles of the PMT dark count and radioactive background, which are evenly distributed in time. Thus, if the contribution to the C.L. from the arrival time distribution is also taken into account, the confidence level will certainly be improved.

\section{Experiment Schedule}\label{Sched}

Referring to Table \ref{table_bkg_afcut}, it is evident that the BRN is the dominant background in this experiment. Therefore, it is crucial to have precise knowledge of the flux, spectrum, and spatial distribution of the BRN on the platform. To achieve this, several liquid scintillator detectors capable of discriminating between neutrons and gamma rays have been installed on the platform to obtain more accurate measurements of the neutron background.

In parallel, the testing of a detector prototype capable of accommodating two $3$ kg CsI crystals is currently underway. The cryogenic system is functioning reliably, and the feasibility of this detector design has been confirmed.

Our plans for 2023 include completing the commissioning of the detector and finalizing the construction of the shielding system. If all goes according to plan, we anticipate starting the data-collection phase in 2024.

\begin{table}[H]
\caption{The event rates in signal region after cut of different kinds of background.
 \label{table_bkg_afcut}}
\newcolumntype{c}{>{\centering\arraybackslash}X}
\begin{tabularx}{\textwidth}{cc}
	\toprule
    \textbf{Background Type} & \textbf{Event Rate in Signal Region after cut/half year/12 kg} \\ \midrule
    Beam related neutron 
      & 666\\ \midrule
    PMT dark count           & 160\\ \midrule
    Radioactive isotopes    & 7\\ \midrule
    Environmental $\gamma$          & negligible\\ \midrule
    Neutrino induced neutron     & negligible\\ \midrule
    Cosmic ray induced radioactive isotopes     & negligible\\ \bottomrule
\end{tabularx}

\end{table}

\section{Summary}

The measurement of CE$\nu$NS signal enjoys great significance among various aspects of physics. When placing a 12 $\text{kg}$ cryogenic undoped CsI detector inside a strong shield on a platform {$7.7\, \text{m}$} away from the Tungsten target in CSNS, with some event selection criteria employed to enhance the signal-to-background ratio, the detector threshold is expected to be lowered to $1.5$$\text{keV}_{\text{nr}}$ (nuclear recoil energy). The detectable CE$\nu$NS event rate is expected to be $160\text{events/half year}$ and the total background rate could be suppressed to $833\text{events/half year}$. Within a half-year data-collection period, a $5\sigma$ detection of CE$\nu$NS signal is anticipated. If the dedicated measurement of neutron background and the test of the prototype progress smoothly, the data collection is anticipated to start in 2 years.

\vspace{6pt} 



\authorcontributions{Conceptualization, Chenguang Su and Qian Liu; methodology, all authors; software, Chenguang Su and Tianjiao Liang; validation, all authors; formal analysis, Chenguang Su and Tianjiao Liang; investigation, all authors; resources, Qian Liu and Tianjiao Liang; data curation, all authors; writing---original draft preparation, Chenguang Su; writing---review and editing, Chenguang Su and Qian Liu; visualization, all authors; supervision, Qian Liu and Tianjiao Linag; project administration, Qian Liu and Tianjiao Liang; funding acquisition, Qian Liu. All authors have read and agreed to the published version of the manuscript.}


\funding{This work was supported by the National Natural Science Foundation of China (Grant No. 12221005 and 12175241) and the Fundamental Research Funds for the Central Universities.}

\institutionalreview{Not applicable.}

\informedconsent{Not applicable.}



\dataavailability{Not applicable.}  

\conflictsofinterest{The authors declare no conflicts of interest.}

\begin{adjustwidth}{-\extralength}{0cm}

\reftitle{References}

\PublishersNote{}
\end{adjustwidth}
\end{document}